\documentclass[12pt]{article}
\usepackage[centertags]{amsmath}
\usepackage{graphicx}

\newcommand{\be}{\begin{equation}}
\newcommand{\ee}{\end{equation}}
\usepackage[centertags]{amsmath}
\usepackage{amsfonts}

\newcommand{\ba}{\begin{eqnarray}}
\newcommand{\ea}{\end{eqnarray}}

\usepackage{amsfonts,amsmath,amssymb,bm}

\begin{document}

\title{\bf  A note on  Chern-Simons CP(1) solitons }
\author{
Lucas Sourrouille$^a$
\\
{\normalsize \it $^a$Departamento de F\'\i sica, FCEyN, Universidad
de Buenos Aires}\\ {\normalsize\it Pab.1, Ciudad Universitaria,
1428, Ciudad de Buenos Aires, Argentina}
\\
{\footnotesize  sourrou@df.uba.ar} } \maketitle

\abstract{We study a gauged $CP(1)$ Chern-Simons system. Using the equations of motion we find, for a static field configuration, that the divergence of the electric field is related to the electric and the $CP(1)$ topological charge densities . We discuss this relation in terms of an ansatz for the fields and use it to show that the $CP(1)$ topological charge is a bound of the energy.   }


\newpage


\vspace{1cm}

 The two dimensional $CP(N)$ sigma model was introduced in the  late  seventies \cite{golo}, in the search of understanding the strong coupling effects in $QCD$. This model captures  several interesting properties, many of them present in four dimensional $QCD$. Both systems share asymptotic
freedom, the generation of a mass gap, confinement, chiral symmetry breaking, anomalies,
soliton solutions and a large $N$ expansion \cite{witten}. Whereas in four dimensional $QCD$ it is difficult to demonstrate the existence of these properties, in two dimensional $CP(N)$ sigma model it becomes comparatively simple. This  relationship between 2d sigma-models and 4d gauge theories has been recently extended in the context of vortex string \cite{tong} showing the existence of a map connecting both theories.

A particular case of $CP(N)$ model is the $CP(1)$ model, or, equivalently, the $O(3)$ sigma model. This model presents topological soliton solutions\cite{polyakov} which can take arbitrary size due to the scale invariance, making them unsuitable as models for particles. One way of breaking the scale invariance, was proposed by Dzyaloshinsky, Polyakov and Wiegmann\cite{polyakov1}. It consists of introducing a Chern-Simons term in the lagrangian and it was suggested in order to study statistics changing properties of that term. In the nineties several works showed that the gauge sigma models can support topological and nontopological soliton solutions\cite{sigmas}, \cite{mehta}. For the particular case of a Chern-Simons-$CP(1)$ system\cite{mehta} was showed, in analogy with occur at the Chern-Simons-Higgs models \cite{jackiw}, that these solitons are static and stable solutions. However, in contrast of what happen with Chern-Simons-Higgs case, these solitons are not self dual solutions, representing it a technical difficult.

In this article we return to the model studied in \cite{mehta}. We are interested in exploring the possibility of finding a static equation for the vortices in that model. We will show that it is possible to find this equation and we will use it to show that the magnetic flux is a multiple of the $CP(1)$ topological charge.

We begin by considering a $(2+1)$ dimensional Chern-Simons model coupled to a complex two component field $n(x)$  described by the action

\begin{eqnarray}
S&=& S_{cs}+\int d^3 x \left(|D_\mu n|^2  - V (n)\right)
\end{eqnarray}

Here  $D_{\mu}= \partial_{\mu} + ieA_{\mu}$ $(\mu =0,1,2)$ is the covariant derivative, $V(n)$ is a positive definite potential and $S_{cs}$ is the Chern-Simons action given by

\begin{eqnarray}
 S_{cs}= \frac{\kappa}{4}\int d^3 x \epsilon^{\mu \nu \rho} A_\mu F_{\nu \rho}= \kappa \int d^3 x \left( A_0 F_{12} + A_2 \partial_0 A_1
\right)
\end{eqnarray}

where
\begin{eqnarray}
F_{\mu \nu}=\partial_{\mu}A_{\nu}-
\partial_{\nu}A_{\mu} \label{F}
\end{eqnarray}

The metric signature is $(1,-1,-1)$ and  the two component field $n(x)$ is subject to
the constraint $n^\dagger n = 1$

Variation of the action yields the field equations

\begin{eqnarray}
D_\mu D^\mu n = -\frac{\partial V}{\partial n^\dagger}
\label{motion1}
\end{eqnarray}
and

\begin{eqnarray}
A_\mu = \frac{1}{2e^2} \left(2ei n^\dagger \partial_\mu n - \frac{\kappa}{2}\epsilon_{\mu \nu \rho} F^{ \nu \rho}\right)
\label{motion}
\end{eqnarray}

The last equation can be rewritten as  $\frac{\kappa}{2}\epsilon_{\mu \nu \rho} F^{ \nu \rho} = - eJ_\mu$ where $J_\mu = -i [n^\dagger D_\mu n - n(D_\mu n)^\dagger]=2\Big(-in^\dagger \partial_\mu n +e A_\mu \Big)$ is the matter current. The time component of Eq. (\ref{motion}),

\begin{eqnarray}
\kappa F_{12} =  -eJ_0 \label{gauss}
\end{eqnarray}

is the Gauss law of the Chern-Simons dynamics. Integrating it over the entire plane one obtain the important consequence that any object with charge $Q =\int d^2 x J_0$ also posses magnetic flux $\Phi = \int B d^2 x$ \cite{Echarge}:

\begin{eqnarray}
\Phi = \frac{-e}{\kappa}Q
\label{Q}
\end{eqnarray}

where in the expression of magnetic flux we renamed $F_{12}$ as $B$.

Using the Gauss law we can write the energy functional for a static field configuration as

\begin{eqnarray}
E=  \int d^2 x \Big(\frac{\kappa^2}{4e^2} B^2 + |D_i n|^2 + V(n) \Big) \,,
\;\;\;\;\;\
i = 1,2  \label{statich}
\end{eqnarray}

Finiteness of the energy requires that as $r \rightarrow \infty$

\begin{eqnarray}
n(\phi)=  \left( \begin{array}{c}
e^{-i\alpha (\phi)}\\
0 \end{array} \right)
\,,
\;\;\;\;\;\
A_\phi (r,\phi)=\frac{1}{e}\nabla_\phi \alpha (\phi)
\label{boundary}
\end{eqnarray}

where  $\alpha$  is a phase angle. The boundary conditions establish a continuous  map between the circle at spatial infinity, $S_1^{phys}$, parameterized by the angle $\phi$ and the internal circle,  $S_1^{int}$, formed by the allowed values of $\alpha (\phi)$.  Therefore finite energy solutions are classified by their winding number $N$ which counts the number of times the phase $\alpha$ winds around the infinite  spatial circle. This implies that $\alpha (\phi) = N \phi$. Then the boundary conditions (\ref{boundary}) become

\begin{eqnarray}
n(\phi)= \left( \begin{array}{c}
e^{-i N \phi}\\
0 \end{array} \right)
\,,
\;\;\;\;\;\
A_\phi (r)= \frac{1}{e}\frac{N}{r}
\label{boundary1}
\end{eqnarray}

Thus the magnetic flux and the electric charge are quantized

\begin{eqnarray}
& \Phi = \int B  d^2 x = \oint A_i dx^i =\frac{2\pi N}{e} \nonumber \\
& Q = \int J_0  d^2 x =-\frac{\kappa}{e^2}2\pi N
\end{eqnarray}

An approach to the boundary conditions can be made by proposing the following ansatz \cite{mehta}:

\begin{eqnarray}
n(\phi, r)=  \left( \begin{array}{c}
\cos(\frac{\theta(r)}{2})e^{-i N \phi}\\
\sin(\frac{\theta(r)}{2} )\end{array} \right)
\,,
\;\;\;\;\;\
 A_\phi (r)= \frac{1}{e}\frac{a(r)}{r}
\label{ansatz}
\end{eqnarray}

Where $\theta(r)$ and $a(r)$ are continuous functions that satisfy

\begin{eqnarray}
\lim_{r \to \infty} \theta(r) = 0
\,,
\;\;\;\;\;\
\lim_{r \to \infty}a(r) = N
\label{13}
\end{eqnarray}

\begin{eqnarray}
\lim_{r \to 0} \theta(r) = \pi
\,,
\;\;\;\;\;\
\lim_{r \to 0} a(r) = 0
\end{eqnarray}

The solutions of the motion equations (\ref{motion1}) and (\ref{motion}) in terms of the ansatz (\ref{ansatz}) were studied for particular choices of the potential term in \cite{mehta}.

Using Eq. (\ref{motion}), we can find an expression for the magnetic field,

\begin{eqnarray}
 B= \epsilon^{ij} \partial_i A_j& =
 \frac{1}{e^2}\Big[\partial_1 (ein^\dagger \partial_2 n) -  \partial_2 (ein^\dagger \partial_1 n)\Big] - \frac{\kappa}{2e^2}\Big[ \partial_1 F_{01} + \partial_2 F_{02}\Big]
 \label{B}
\end{eqnarray}

The first two terms can be rewritten taking into account that $ \epsilon^{i j}\partial_i (ein^\dagger \partial_j n) = ei\epsilon^{i j} (D_i n)^\dagger (D^j n)$, while the third and fourth terms form the divergence of the electric field. When the fields are in a static configuration, $\frac{ei}{2\pi}\epsilon^{i j} (D_i n)^\dagger (D^j n)$ is the $CP(1)$ topological charge density $J^{CP(1)}_0$ (see Ref. \cite{witten}), so in that case the equation (\ref{B}) can be written as:

\begin{eqnarray}
B =  \frac{2\pi}{e^2} J^{CP(1)}_0 - \frac{\kappa}{2e^2}\nabla . {\bf E} \label{B2}
\end{eqnarray}

Using Eq.(\ref{gauss}) we can write:

\begin{eqnarray}
\nabla . {\bf E} = \frac{2e^3}{\kappa^2}J_0 + \frac{4\pi}{\kappa} J^{CP(1)}_0
\label{Div}
\end{eqnarray}
Here, in contrast with the Maxwell theory, the electric monopole is generated by both electric  and $CP(1)$ topological charge densities. Clearly the solutions of Eq.(\ref{Div}) are the static solutions of Eq.(\ref{motion}). In terms of the ansatz (\ref{ansatz}) Eq. (\ref{Div}) reads

\begin{eqnarray}
\nabla . {\bf E} =\frac{1}{r}\partial_r (rE_r)= -\frac{2e}{\kappa}\frac{\partial_r a(r)}{r} - \frac{2e}{\kappa} \frac{N}{r}\cos(\frac{\theta(r)}{2})\sin(\frac{\theta(r)}{2} ) \partial_r \theta(r)
\label{D1}
\end{eqnarray}
Note that $J^{CP(1)}_0 \not= 0$ implies no trivial solution for Eq.(\ref{D1}) and so $J_0 \not=0$. Conversely  $J_0 \not=0$ implies $N\not=0$ and then  $J^{CP(1)}_0 \not= 0$. When  $J^{CP(1)}_0 = 0$, $J_0 =0$ and the  converse is also true. So we see that the existence of one current implies the existence of the other one.  Also note from (\ref{D1}) that there exists  a circle of radius $r_1$ for which $\cos(\frac{\theta(r_1)}{2})=0$. In that region the electric field is generated only by electric charge.

We can obtain an expression for the magnetic flux, by integrating (\ref{B2})

\begin{eqnarray}
\Phi = \int B d^2 x = \frac{2 \pi}{e^2} Q_{CP(1)} -\frac{\kappa}{2e^2} \int d^2 x \nabla . {\bf E}
\label{M}
\end{eqnarray}
where $Q_{CP(1)}$ is the $CP(1)$ topological charge. The integral of the electric field divergence can be evaluated with the help of the relation $\nabla . {\bf E} = \nabla \times {\bf J}$, which follows from Eq.(\ref{motion}). Integrating it and using the Stokes theorem we have $\int d^2 x \nabla . {\bf E} = \oint J_j dx^j$, where line integral is over the infinity circle. Replacing the ansatz (\ref{ansatz}) in the expression of  $J_\phi$, we find $J_\phi = \Big(-a(r) + \cos^2(\frac{\theta(r)}{2})N\Big)\frac{2e}{r\kappa}$ which, in virtue of (\ref{13}), goes to zero faster than $\frac{1}{r}$ when  $r \rightarrow \infty$.
Thus $\int d^2 x \nabla . {\bf E}= 0$ and  Eq.(\ref{M}) for the magnetic flux reduces to

\begin{eqnarray}
\Phi = \int B d^2 x =  \frac{2 \pi}{e^2}  Q_{CP(1)}
\label{P}
\end{eqnarray}
So, the magnetic flux and also the electric charge are multiples of  $CP(1)$ topological charge. Note we can deduce, from (\ref{Div}), the quantization of the electric charge with the only requirement that the electric field satisfied $\int d^2 x \nabla . {\bf E}= 0$. We can use Eq.(\ref{P}) to show that $\frac{2\pi}{e^2} Q_{CP(1)}$ is a topological bound of functional energy (\ref{statich}). Using the fundamental relation $|D_i n|^2 =   |( D_1 \pm iD_2)n|^2 \mp B \pm \epsilon^{ij} \partial_i J_j$, Eq. (\ref{statich}) changes to:

\begin{eqnarray}
E =\int d^2 x \left( \frac{\kappa^2}{4e^2} B^2 +  |( D_1 \pm iD_2)n|^2 \mp B \pm \epsilon^{ij} \partial_i J_j + V(n)\right)
\label{H}
\end{eqnarray}
In virtue of the above discussion we can write

\begin{eqnarray}
E=\int d^2 x \Big( \frac{\kappa^2}{4e^2} B^2 +  |( D_1 \pm iD_2)n|^2   \mp \frac{2\pi}{e^2} J_{CP(1)}^0 + V(n)\Big)
\label{H2}
\end{eqnarray}
which implies:

\begin{eqnarray}
E \geq \frac{ 2\pi}{e^2} \int d^2 x J_{CP(1)}^0 =  \frac{2\pi}{e^2} Q_{CP(1)}
\label{L}
\end{eqnarray}

Finally, it is interesting to see what happens when the ansatz (\ref{ansatz}) change to

\begin{eqnarray}
n(\phi, r)=  \left( \begin{array}{c}
\cos(\frac{\theta(r)}{2})e^{-i N \phi}\\
\sin(\frac{\theta(r)}{2} )\end{array} \right)
\,,
\;\;\;\;\;\
A_\phi (r)= \frac{1}{e}\left(\frac{a(r) + constant}{r} \right)
\label{ansatz1}
\end{eqnarray}

where with constant we mean a real number distinct from zero. In that case the functional energy (\ref{statich})
is not bounded in infinite region of integration. After some algebra the equation (\ref{D1}) can be rewritten in terms of the new
ansatz (\ref{ansatz1}) as

\begin{eqnarray}
\frac{\kappa^2}{4e^4}\Big(- \frac{\partial_r a(r)}{r^2} + \frac{\partial_r^2 a(r)}{r} \Big) + \frac{a(r)+ constant}{r}-\cos^2 (\frac{\theta (r)}{2}) \frac{N}{r}= 0
\label{ansatz2}
\end{eqnarray}

This is an inhomogeneous equation even when $N=0$. So if the solution exists for $N=0$, we have a solution that
satisfies the Gauss law of Maxwell theory, that is $\nabla . {\bf E} = \frac{2e^3}{\kappa^2}J_0$.

\vspace{2cm}
In this letter we have showed that the Euler-Lagrange equations can be combined in a way that, for a static field configuration, the divergence of electric field is related to electric and topological charge densities. Using this result we have noted that the existence of one density implies the existence of the other one and that the $CP(1)$ topological charge is a bound of the energy. We think that this result  may be useful in the understanding of  the  Cherm-Simons sigma models.

\vspace{2cm}
{\bf Acknowledgements}

I would like to thank Jer\'{o}nimo Peralta Ramos for comments and useful discussions, Oliver Piguet for encouragement and comments and Wojtek Zakrzewski, Gustavo Lozano, Bernard Piette, Gaston Giribet, Guillermo Silva and Diego Marques for comments.

.

\end{document}